\documentclass[%
reprint,superscriptaddress,
 amsmath,amssymb,
 aps,
prb,
]{revtex4-2}

\usepackage{overpic}
\usepackage{graphicx}
\usepackage{dcolumn}
\usepackage{bm}
\usepackage[normalem]{ulem}
\usepackage{color}

\begin{document}
\title{Anharmonicity and Charge-Noise Sensitivity of Fraunhofer Qubit}
\author{Longyu Ma} 
\author{Tony Liu} 
\affiliation{Department of Physics, University of Wisconsin–Madison, Madison, Wisconsin 53562, USA}
\author{Javad Shabani}
\affiliation{Center for Quantum Information Physics, Department of Physics, New York University, New York 10003, USA}
\author{Kasra Sardashti}
\affiliation{Department of Physics, University of Maryland, College Park, Maryland 20740, USA}
\affiliation{Laboratory for Physical Sciences, University of Maryland, College Park, Maryland 20740, USA}
\author{Vladimir E. Manucharyan}
\affiliation{Institute of Physics, Ecole Polytechnique Federale de Lausanne, CH 1015}
\author{Maxim G. Vavilov}
\affiliation{Department of Physics, University of Wisconsin–Madison, Madison, Wisconsin 53562, USA}

\begin{abstract}
We present a theory of a flux-tunable superconducting qubit, the "Fraunhofer qubit," based on the Fraunhofer interference in a wide ballistic Josephson junction. As magnetic flux threads the junction, the Josephson potential is effectively averaged over a phase window proportional to $\Phi$.
For perfectly transmitting junctions, as flux approaches one flux quantum $h/2e$, the flux averaging transforms the potential near its minimum from a quadratic to a triangular shape, resulting in significantly enhanced anharmonicity. 
This enhancement persists for junctions with lower transparency conducting channels. 
Microscopic tight-binding simulations that include inhomogeneous electrostatic potential and disorder confirm the enhancement of anharmonicity.
These results establish a framework for flux control in hybrid superconducting circuits, providing an operating point where anharmonicity and charge-noise protection can be optimally balanced.

\end{abstract}

\date{\today}
\maketitle

\emph{Introduction.} The transmon is a leading qubit for quantum information processing based on superconducting circuits due to its relatively simple design and insensitivity to charge noise \cite{Koch_2007,Clarke,DiVincenzo_2000, PhysRevB.77.180502,PhysRevLett.101.080502}. These devices feature capacitively shunted Josephson junctions; the latter act as nonlinear inductors, resulting in a weakly anharmonic transmon energy spectrum.
By replacing the single Josephson junction with a superconducting quantum interference device (SQUID), the energy levels can be modified by passing a magnetic flux through the SQUID \cite{PhysRevLett.111.080502}. Although this design allows for flux tunability, the SQUID loops introduce an additional sensitivity to flux noise.

Alternatively, the qubit can be controlled by applying a gate voltage to the junction, forming a "gatemon" \cite{Casparis_2016,PhysRevLett.115.127001, de_Lange_2015,Pita_Vidal_2020, Casparis_2018,zhuo2023hole, zheng2024coherent,sagi2024gate}. The gate voltage modulates the electron density in the superconductor-semiconductor-superconductor junction, thereby tuning the Josephson energy and making the qubit frequency tunable. In practice, however, this tunability comes at a cost: gatemons often exhibit increased decoherence \cite{PRXQuantum.4.030339,PhysRevResearch.6.023094}, as the gate channel introduces additional loss mechanisms and fluctuating electrostatic environments that couple to the qubit. Moreover, high-transmission semiconductor weak links tend to reduce the qubit anharmonicity \cite{PhysRevB.97.060508}, limiting the spectral separation between computational and higher levels and therefore constraining gate speeds and control fidelity. These two effects underscore the main trade-off of gate-based operation. While voltage control offers fast, dissipationless tunability, it can simultaneously diminish the qubit's coherence and anharmonicity, posing challenges for the realization of high-fidelity gates.

In this paper, we characterize a "Fraunhofer qubit" (Fig. 1) that utilizes Fraunhofer interference in a single Josephson junction to achieve flux tunability instead of using a SQUID. By applying a perpendicular magnetic field to the junction, the Josephson potential undergoes oscillations with respect to the magnetic flux, mirroring the Fraunhofer patterns typically observed in critical current \cite{PhysRevLett.11.200, AkhermovHourglass, AkhmerovBeenakker, antsygina1975josephson, Glazman}.
Our primary observation is that at magnetic flux levels below the superconducting flux quantum, the qubit's anharmonicity is significantly enhanced. At the same time, the sensitivity to charge noise remains exponentially suppressed, preserving the key operational advantage of the transmon architecture. 
This behavior is consistent with recent findings in flux-frustrated gatemons \cite{Liu.2025}, where the interference of higher-order harmonics of the supercurrent leads to enhanced anharmonicity, particularly at half-integer flux sweet-spots. It is important to emphasize that the Fraunhofer qubit achieves this tunability and enhances anharmonicity without requiring a SQUID formed by high-inductance superconducting wires.

\begin{figure}[htb]
\label{fig1}
  \centering
  \includegraphics[width=\linewidth]{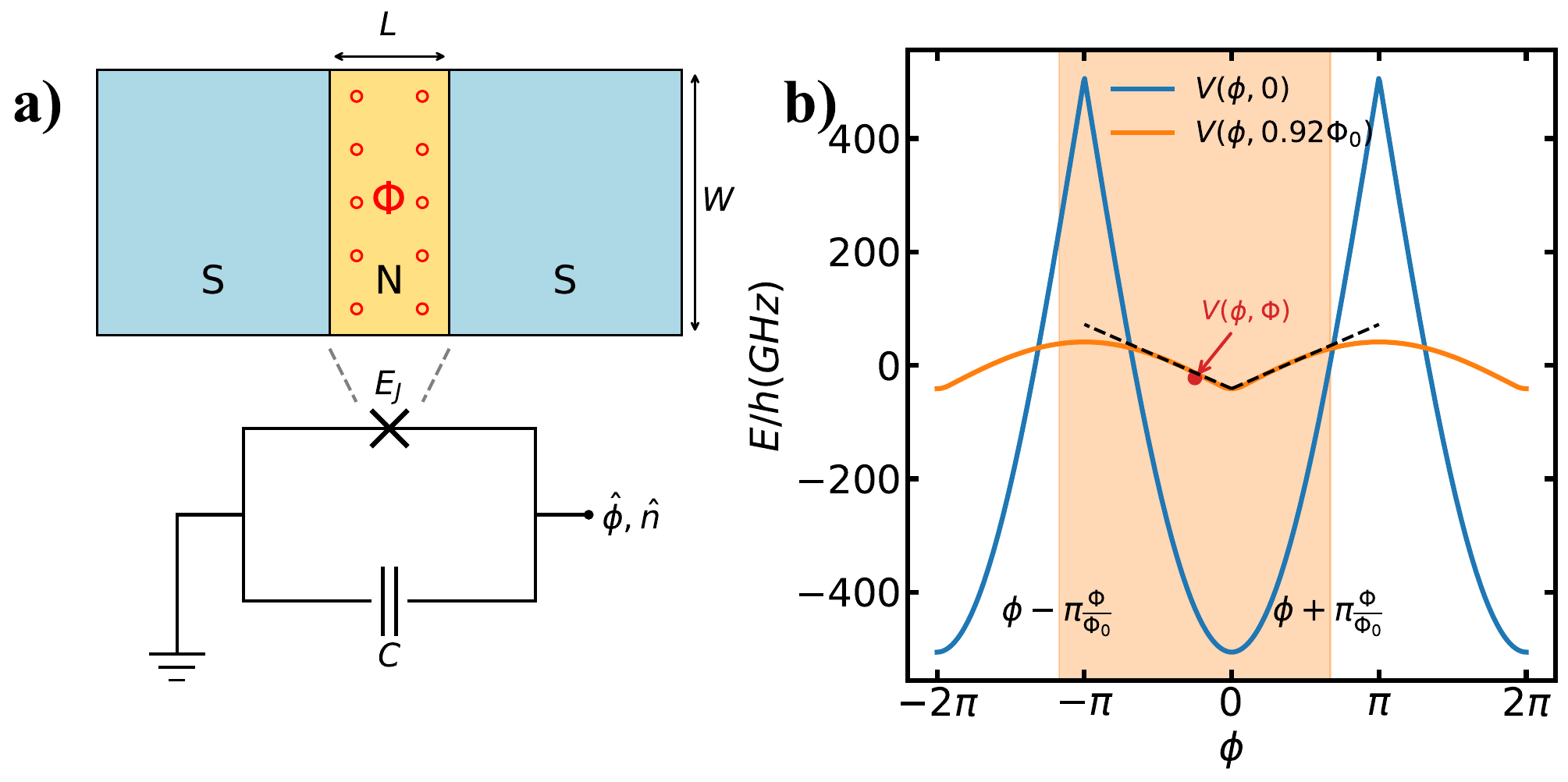}
  \caption{
  Schematic of the Fraunhofer qubit. (a) Circuit diagram of the Fraunhofer qubit with zoom-in view of the Josephson junction, illustrating the external magnetic field applied in the normal region. (b) Illustration of the flux averaging window for potential, with black dashed lines for the triangular potential demonstration.
  }
  \label{SNSfraunhaufer}
\end{figure}

\begin{figure}[htb]
\label{fig2}
      \centering
    \includegraphics[width=0.95\linewidth]{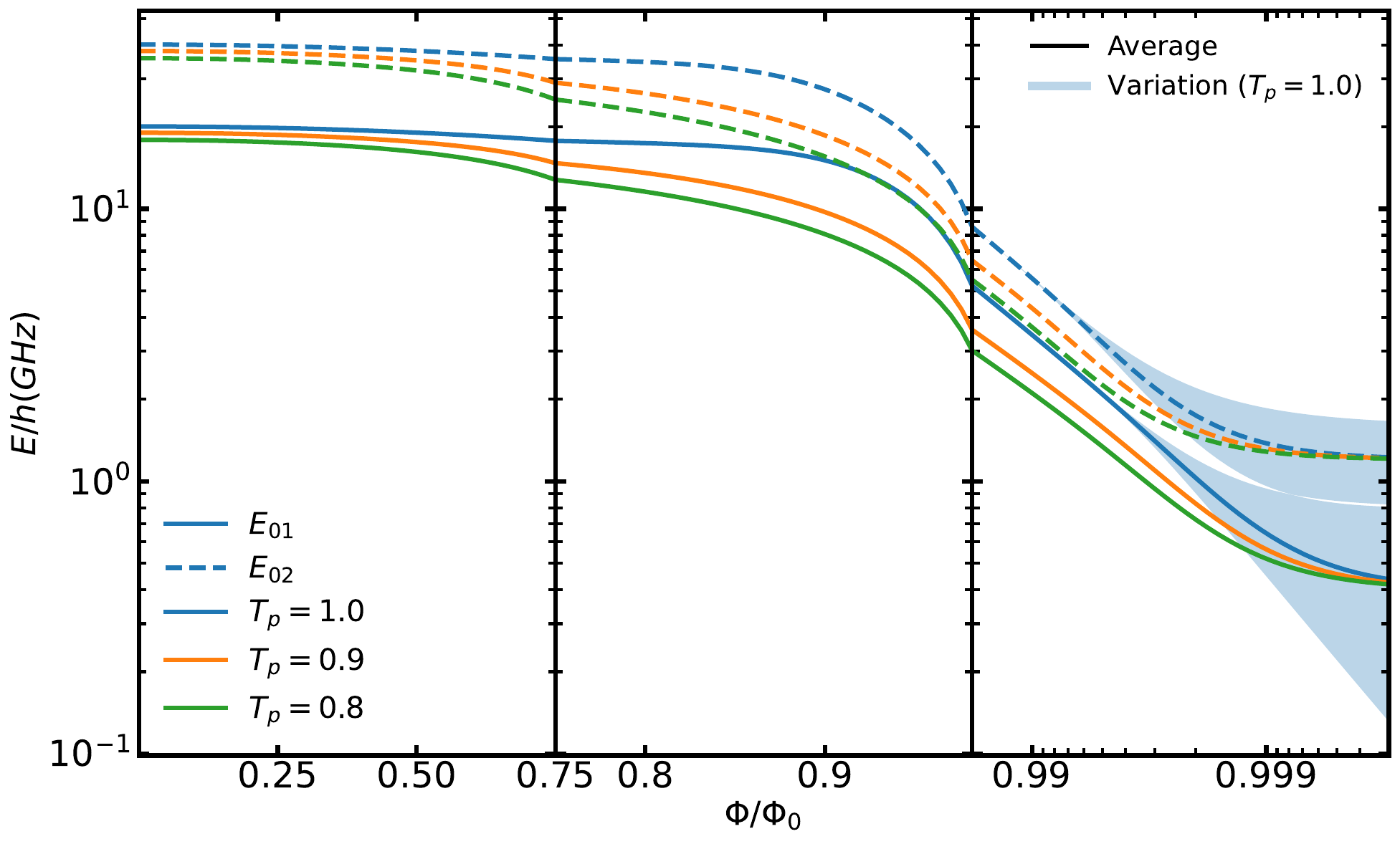}%
  \caption{
Frequencies for $0\to1$ and $0\to 2$ qubit transitions as functions of magnetic flux, shown for different junction transparencies. In the low-flux regime, the perfect junction case exhibits an essentially flux-independent transition frequency, while junctions with lower transparencies show flux tunability at lower $\Phi$. 
In the intermediate-flux range (the middle panel), the qubit frequencies change faster with the flux for all values of transparency $T$.
At high flux (the right panel), the transition frequencies are strongly suppressed, and the qubit gradually enters a charge-sensitive regime of a Cooper-pair box. 
Computations were performed for $E_C/h=200$MHz, $\Delta/h=50.7$ GHz and $N=20$ channels.}
\end{figure}

\emph{Model.} We start with an analytical expression for the Josephson potential based on the Beenakker\cite{PhysRevLett.67.3836, AkhmerovBeenakker,RevModPhys.69.731} formula for Andreev bound state energies.
In the case of a wide junction with perfect interfaces, the magnetic field effectively averages the Andreev bound states over a phase interval proportional to the flux through the junction. As the flux increases from zero toward the superconducting flux quantum, the effective Josephson energy decreases. Simultaneously, the shape of the potential near its minimum transitions from a weakly anharmonic well to a triangular well, highlighted by the dashed lines in Fig. 1b, which significantly enhances the anharmonicity. We find a regime in which the potential becomes triangular as the ratio of Josephson energy to charging energy is large, thereby allowing increased anharmonicity even in the deep-well limit. Numerical simulations show that these features persist even in junctions with imperfect interfaces.

The Hamiltonian for the qubit includes both charging and Josephson terms:
\begin{equation}  
\hat H = 4E_{C}(\hat n - n_{g})^{2} + V(\hat \phi,\Phi).
\label{eq:FullHamiltonian}
\end{equation}
Here $\hat n$ is the number operator for the electrons in the junction and $\hat \phi$ is the operator for the phase difference across the junction, obeying the commutation relation $[\hat{\phi}, \hat{n}] = i$. $E_{C}$ is the charging energy, and $n_{g}$ is the offset charge. 
\begin{figure}[htb]
    \centering
    % 1. Note: overpic REPLACES \includegraphics. 
    % Put the [width] options in the square brackets here.
    \begin{overpic}[width=0.95\linewidth]{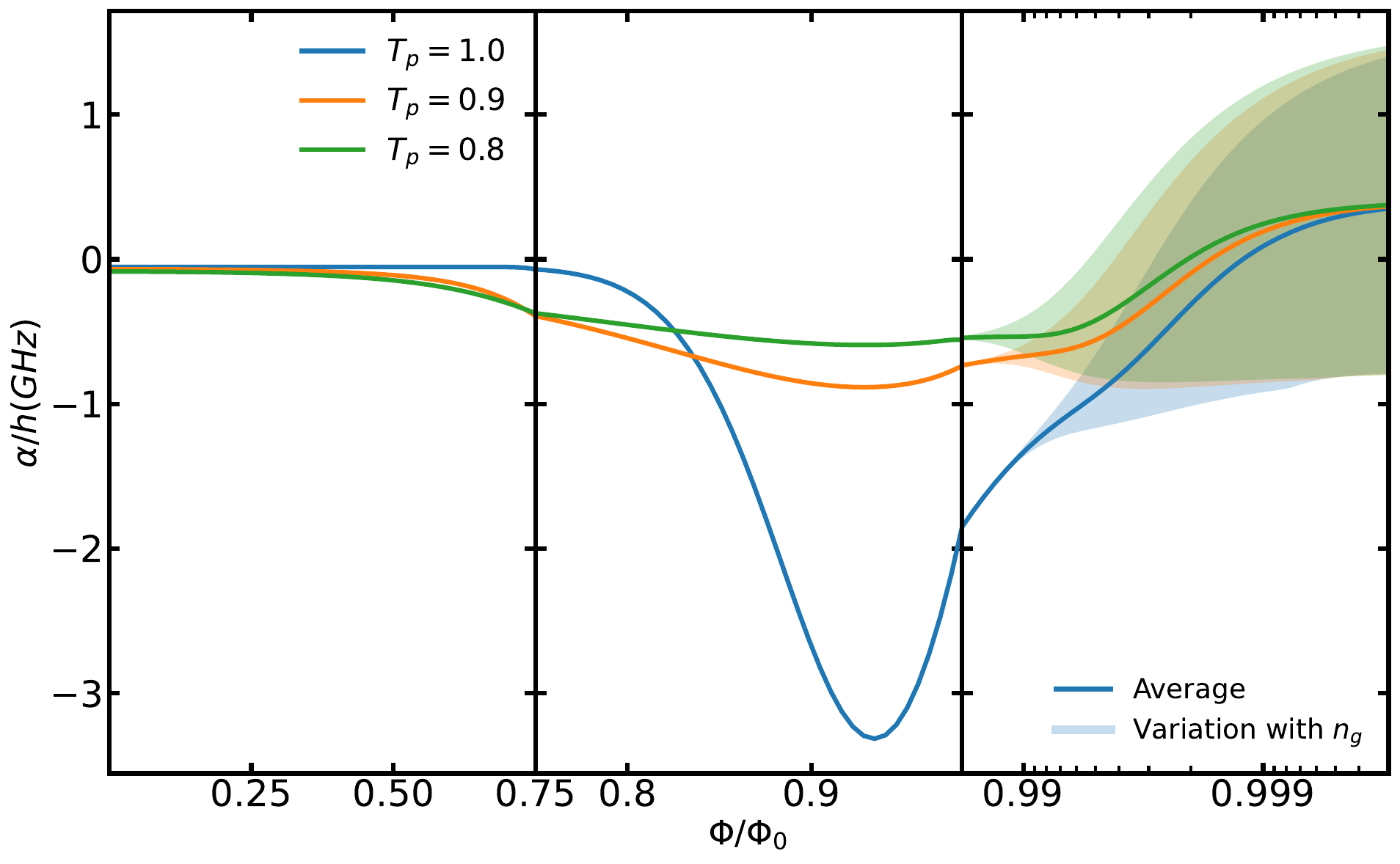}
    
        % 2. Overlay the inset
        \put(8.5, 7){\includegraphics[width=0.37\linewidth]{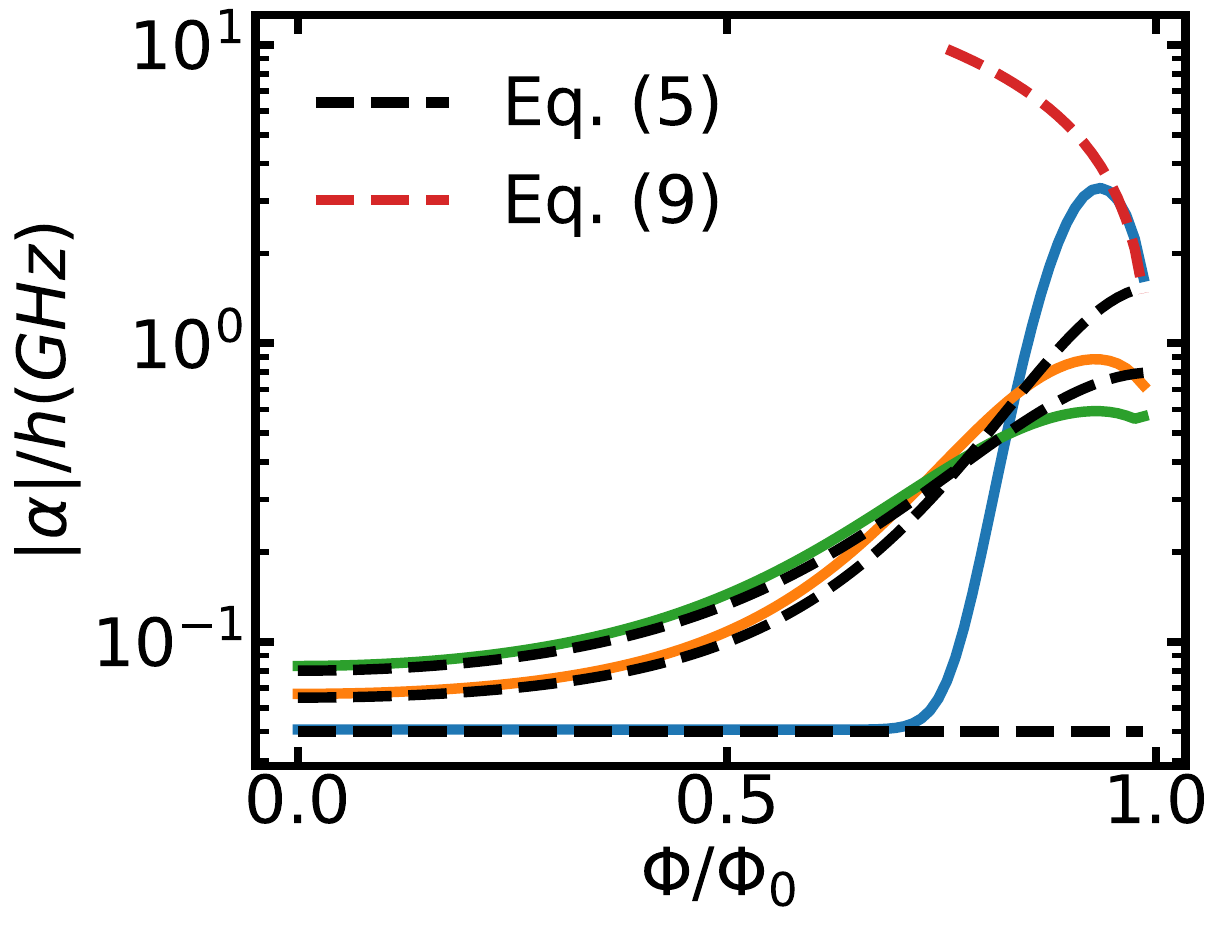}}
        
    \end{overpic}
\caption{Anharmonicity (Eq.~\eqref{eq:anharmonicitydef}) of the qubit as a function of magnetic flux for the same parameters as in Fig.~2
In the low-flux regime, the anharmonicity of a reflectionless junction is flux-independent and remains at $E_C/4$ as highlighted in the logarithmic-scale inset. The inset compares the anharmonicity versus flux with perturbative and triangular-well analytic predictions, with colors matched to those used in the main panel.
In the intermediate-flux range, the reflectionless junction exhibits pronounced anharmonicity enhancement, which broadens as channel transparency decreases.
At high flux, the potential well becomes increasingly shallow, and the qubit crosses over into a charge-sensitive regime.}

\end{figure}
The potential term $V(\phi, \Phi)$ describes the energy of a wide rectangular junction where the width is much larger than the length. In this regime, a uniform perpendicular magnetic field effectively averages the zero-field potential over a phase window $2\pi \Phi/\Phi_0$ (See Appendix A):
\begin{equation}
V(\phi, \Phi) = \sum_{p=1}^N \frac{\Phi_0}{2 \pi \Phi} \int^{\phi + \pi \Phi/\Phi_0}_{\phi - \pi \Phi/\Phi_0} d \theta E_p(\theta).
\label{fluxAverage}
\end{equation}
In this expression, the modified energies are calculated using a semiclassical method in which bound states are associated with semiclassical electron-hole trajectories \cite{AkhermovHourglass}. We consider a single ballistic Josephson junction in the short-junction limit, where the length of the junction is much smaller than the superconducting coherence length $\hbar v_F/\Delta$, $v_F$ and $\Delta$ are the Fermi velocity and superconducting gap. In this case, each transverse channel $p$ supports a single Andreev boundstate. Following the work of Akhmerov and Beenakker \cite{PhysRevLett.67.3836, AkhmerovBeenakker,RevModPhys.69.731} , these zero-field energies $E_p(\phi)$ are defined as:
\begin{equation}
E_{p}(\phi) = \Delta \sqrt{1 - T_{p} \sin^{2}\left(\frac{\phi}{2}\right)}.
\label{eq:AndreevEnergy}
\end{equation}
Here $\Delta$ is the superconducting gap in the electrodes, and $T_{p} \in [0,1]$ is the transmission probability of channel $p$.

\emph{Anharmonicity.} We use a conventional charging energy $E_C/h = 200\,\mathrm{MHz}$, a superconducting gap $\Delta/h = 50.7\,\mathrm{GHz}$ \cite{PRXQuantum.4.030339,PhysRevLett.126.036802}, and $N = 20$ channels.  The full Hamiltonian, Eq.~\eqref{eq:FullHamiltonian} 
is solved numerically by discretizing in a phase basis, as described in Appendix~B. The resulting transition energies between states $0\to 1$ ($0\to 2$) exhibit tunability with magnetic flux as shown in Fig.~2.

An important characteristic of the transmon-like qubits is  their anharmonicity $\alpha$:
\begin{equation} 
\alpha = E_2 - 2E_1 + E_0.
\label{eq:anharmonicitydef}
\end{equation}
We observe a significant enhancement of the anharmonicity at the moderate flux values $\Phi \lesssim \Phi_0$, see Fig.~3.

To investigate this behavior of the anharmonicity, we expand the potential $V(\phi,\Phi)$ given by Eq.~\eqref{fluxAverage} to the quartic order in $\phi$.
Carrying out the perturbation theory analysis for a junction characterized by a single transmission coefficient $T_p = T$, we find that the anharmonicity is
\begin{align}
\alpha_{\mathrm{pert}}(\Phi) 
= -\frac{E_{C}}{4}\left(1
+ \frac{3\,(1-T)}{\left[\,1 - T\,\sin^{2}\!\bigl(\pi\,\Phi/(2\Phi_{0})\bigr)\right]^2}\right).
\label{eq:anharmonicityPert}
\end{align}
From Eq.~\eqref{eq:anharmonicityPert}, we observe that for $T < 1$, 
the magnitude of the anharmonicity increases with $\Phi$, see the inset in Fig.~3.
The result obtained from numerical solution of the Hamiltonian, Eq.~\eqref{eq:FullHamiltonian},  is in good agreement with the perturbative result, Eq.~\eqref{eq:anharmonicityPert},  for typical values of $T<1$ and  $\Phi/\Phi_0 \lesssim 1$. As $\Phi\to \Phi_0$, the Josephson energy is eventually reduced to the point where the qubit is highly charge sensitive, and the perturbative expansion of the potential in the deep well limit is no longer appropriate.

\emph{Reflectionless channels. } The case $T =1$ is different. Equation~\eqref{eq:anharmonicityPert} gives  flux-independent $\alpha_{\mathrm{pert}} = -E_{C}/4$. While this is consistent with the numerics at small values of $\Phi$ (see Fig.~3), the numerical results indicate that $\alpha$ begins to increase at $\Phi/\Phi_0 \approx 0.75$.
This discrepancy develops when a weakly anharmonic approximation, Eq.~\eqref{eq:anharmonicityPert}, breaks down as $\Phi \rightarrow \Phi_0$. 
As $\phi$ continues to increase, the right side of the phase averaging window crosses the cusp located at $\phi = \pi$  and the potential $V(\phi, \Phi)$ transitions to a triangular shape (see Fig. 1b).
To demonstrate this crossover, we can explicitly expand Eq.~\eqref{fluxAverage} for $\phi \ll 1$, and obtain
\begin{align}
    V(\phi, \Phi) -  V(0, \Phi)   
\approx \widetilde {E_J}(\Phi)
    \begin{cases}
        (1/2)  \phi^2/\phi_T, &  |\phi| <\phi_T \\
          |\phi| -\phi_T/2,
        & |\phi| > \phi_T
    \end{cases}
\end{align}
where the "detuning width", $ \phi_T = \pi \left(1 - \Phi / \Phi_0\right)$  describes the size of the phase interval near $\phi = 0$ where the potential is quadratic and
\begin{align}
\widetilde {E_J}(\Phi)= -\sum_p\frac{\Delta}{\pi}\,\frac{\Phi_{0}}{\Phi}
   \Bigl[\,1 - \cos\!\Bigl(\tfrac{\pi\Phi}{2\Phi_{0}}\Bigr)
            - \sin\!\Bigl(\tfrac{\pi\Phi}{2\Phi_{0}}\Bigr)
   \Bigr].
\end{align}
is the effective Josephson energy $\widetilde {E_J}(\Phi)=[V(\pi, \Phi) -  V(0, \Phi)]/2 $ reduced by the magnetic flux through the junction. 
At small fields where $\Phi/\Phi_0 \ll 1$, the detuning width is of order 1 and it is appropriate to use a weakly anharmonic well approximation.
However, in the limit $\phi_T \rightarrow 0$, the quadratic portion of the potential collapses, and the potential should instead be approximated by a triangular well.
The energy levels corresponding to this triangular potential can then be obtained by solving the Schrödinger equation for a piecewise-linear potential, which reduces to the Airy differential equation.
The energy levels in this limit are then given by,
\begin{equation}
E_n(\Phi)
= 4^{1/3} x_n\, E_C
\left( \frac{\widetilde{E}_J(\Phi)}{E_C} \right)^{2/3}.
\end{equation}
where \(x_{n}\) denotes the relevant Airy zeros. 
For the lowest three levels, we take
\(x_{0} \approx 1.0188\), \(x_{1} \approx 2.3381\), and \(x_{2} \approx 3.2482\), corresponding to the first zeros of the Airy function derivative (even states) and the Airy function (odd states), respectively. 
The resulting anharmonicity is then given in terms of $\eta = 4^{1/3}(x_2-2x_1+x_0)$ as
\begin{equation}
    \alpha_{\mathrm{tri}}(\Phi)
    = \eta  E_{C}\left( \frac{\widetilde {E_J}(\Phi) }{E_C}\right)^{2/3},\quad \eta \approx -0.6496.
    \label{eq:triangularAnharm}
\end{equation}
Note that $\alpha_{\mathrm{tri}}$ is much larger than the flat anharmonicity in the low field regime.
Comparing $\alpha_{\mathrm{tri}}$ with the numeric results, we see that there is good agreement when $\Phi$ is close to $\Phi_0$ but still not the Cooper pair box.

The nonmonotonic behavior of $\alpha_{\text{tri}}(\Phi)$ follows from the evolution of the potential with flux. As $\Phi$ approaches $\Phi_{0}$, the well becomes increasingly triangular, which enhances the anharmonicity. Beyond this point, the potential shape remains essentially triangular while the overall Josephson scale $\widetilde{E}_{J}(\Phi)$ continues to decrease, causing anharmonicity to drop. In contrast to the conventional transmon, the anharmonicity of the Fraunhofer qubit is enhanced by the additional scaling factor $\alpha_{\text{tri}} \propto (\widetilde{E}_{J}/E_{C})^{2/3}$, when increasing $\widetilde{E}_{J}$ increases the anharmonicity.

As $\Phi\to \Phi_0$, the depth of the Josephson potential, $\widetilde {E_J}$, decreases, making the qubit sensitive to the offset charge fluctuations, $n_g$ in Hamiltonian~\eqref{eq:FullHamiltonian}. 
The charge dispersion for the $m$-th energy level can  be approximated in the form for the transmon qubits with the flux-dependent $\widetilde {E_J} = \widetilde {E_J}(\Phi)$:
\begin{equation}
\epsilon_m(\Phi) \simeq 
\frac{2^{4m+5}}{m!} \sqrt{\frac{2}{\pi}}
 E_C 
\left(\frac{\widetilde {E_J}}{2E_C}\right)^{\frac{m}{2} + \frac{3}{4}}
\exp\left(-\sqrt{\frac{8\widetilde {E_J}}{E_C}}\right).
\end{equation}
The Fraunhofer qubit energy spectrum remains exponentially insensitive to the charge noise provided that $\widetilde {E_J}(\Phi) \gg E_C$.

\begin{figure}[!]
\centering
\includegraphics[width=\linewidth]{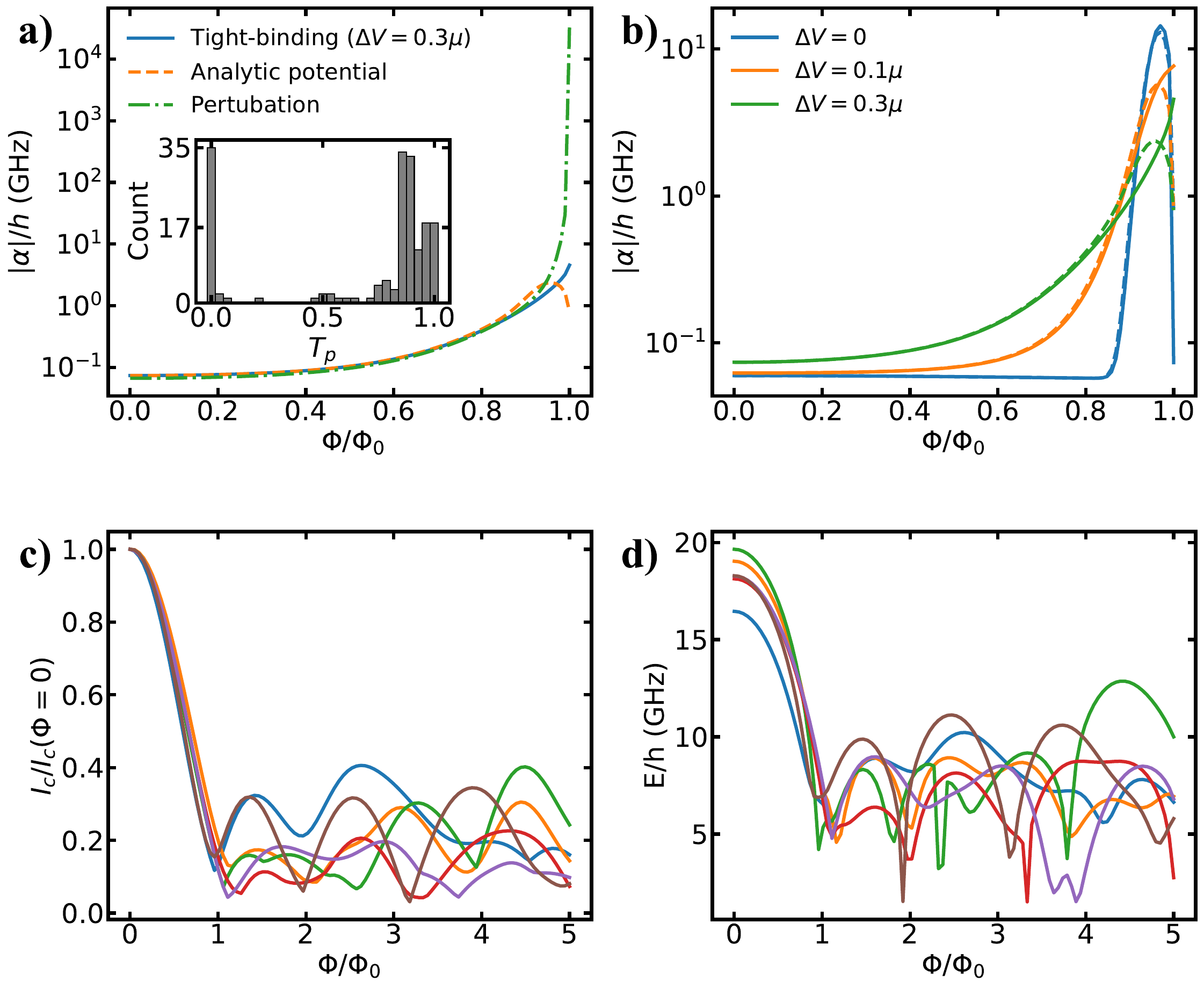}%
\caption{Tight-binding simulation. (a) Comparison of anharmonicity obtained from various approaches for $\Delta V = 0.3\mu$. The inset shows the transparency distribution extracted from the tight-binding model in the absence of magnetic flux. (b) Anharmonicity versus flux for different values of $\Delta V$. Solid lines represent results from the tight-binding model, while dashed lines correspond to calculations using the analytic potential, with tight-binding model transparencies. (c) Critical current and (d) qubit frequency as functions of flux for several disorder realizations.}
  \label{fig4}
\end{figure}

\emph{Inhomogeneous electrostatic potential.} To validate our analytical framework, we perform microscopic simulations using a tight-binding model (see Appendix C and Ref.~\cite{Groth_2014}). By incorporating a potential well in the normal region, we generate a distribution of transmission coefficients $\{T_p\}$. Using this numerical setup, we compute the flux-dependent anharmonicity and compare it with the predictions of Eq.~\eqref{fluxAverage}, where the transmission distribution is extracted from the same model in the zero-flux limit.

The presence of a distribution of transmission channels introduces more intricate behavior. The triangular potential approximation remains valid primarily for highly transparent channels where standard perturbation theory breaks down. Conversely, channels with lower transparency align with the perturbative predictions of Eq.~\eqref{eq:anharmonicityPert}, yielding a broader and weaker anharmonicity peak. While the overall system response is a composite that cannot be fully captured by either limit alone, the tight-binding simulations consistently confirm the predicted enhancement of anharmonicity, showing strong agreement with our analytical results.

\emph{Disorder. } We also investigated the effect of disorder on the qubit energy spectrum through numerical simulations. To model this behavior, we employed a tight-binding framework in which disorder is introduced by a random shift of the chemical potential within the normal region. In our analysis, the mean free path $l_0=13$ nm is shorter than the junction length scale $L=100$ nm. In this regime, the dependence of the critical current on the magnetic flux remains nearly universal for flux values below the flux quantum $\Phi < \Phi_0$, as shown in Fig.~4c. However, as the flux exceeds $\Phi_0$, the critical current exhibits irregular variations that are unique to each specific disorder realization, acting as "magnetic fingerprints." A similar behavior develops in the qubit transition frequency: the frequency decreases monotonically as the flux increases from zero to $\Phi_0$, but it shows an irregular dependence for $\Phi > \Phi_0$. In this high-flux regime, the qubit displays a sequence of local frequency maxima and minima that can be utilized as first-order flux-insensitive sweet spots for qubit operation.

The clean limit offers flux-insensitive sweet spots that exhibit significantly reduced protection against charge noise. In contrast, the presence of disorder prevents the magnetic field from strongly suppressing the barrier height $\widetilde{E_J}(\Phi)$, giving rise to multiple "sweep spots" where the qubit frequency remains within a practically usable range and insensitive to charge noise. This observation suggests that disordered junctions can confer a strategic advantage by creating additional flux sweet spots without sacrificing the system's necessary charge protection.

\emph{Conclusions. } We have characterized the "Fraunhofer qubit," a device in which a uniform perpendicular magnetic field serves as a tool to reshape the qubit's Josephson energy landscape. We demonstrated that the magnetic field provides a controllable trade-off between anharmonicity and charge-noise sensitivity. This analytical framework, rooted in the short-junction formula for Andreev bound states \cite{PhysRevLett.67.3836,RevModPhys.69.731}, reveals how phase variation in wide junctions enhances anharmonicity while simultaneously tuning the qubit frequency.

Our findings are further substantiated by microscopic tight-binding simulations, which verify that the macroscopic potential $V(\phi, \Phi)$ accurately captures the collective behavior of diverse transmission channels $\{T_p\}$. This alignment between analytics and numerics establishes a coherent picture of flux-controlled hybrid qubits. Future investigations incorporating material defects and spin–orbit coupling will further refine the modeling of these devices, paving the way for optimized gatemon architectures.

\emph{Acknowledgments.} We thank Arunav Bordoloi, Shukai Liu, and Charles Marcus for stimulating discussions. This work has received funding from the ARO NextNEQST program (contract no. W911-NF22-10048).
Some parts of our numerical simulations were performed using the kwant~\cite{Groth_2014} and QuTiP~\cite{JOHANSSON20121760, JOHANSSON20131234} python packages.

\bibliography{Fqubit_refs}{}
\newpage
\appendix
\section{Semiclassical calculation of the Josephson potential}
\begin{figure}[htb]
  \centering
  \includegraphics[width=\linewidth]{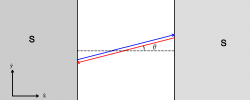}%
  \caption{An example of a quasi-classical electron/hole trajectory which contributes to the bound state energy.}
  \label{SNStrajectories}
\end{figure}

For a ballistic SNS junction, the Josephson potential can be obtained in the semiclassical approximation \cite{AkhmerovBeenakker,AkhermovHourglass}, where boundstates of the junction are associated with periodic electron/hole trajectories that reflect off the NS boundaries. We consider the low-field regime, in which the cyclotron radius is small compared with the junction size and the classical trajectories are straight lines between the superconducting electrodes, as shown in Fig.~\ref{SNStrajectories}.
Each trajectory can be uniquely labeled by the pair of parameters $(y, \theta)$, where $y$ is the y-coordinate of the midpoint and $\theta$ is the angle formed with the x-axis.
The boundstate of a particular transverse mode $p$ is associated with all trajectories with angle $\theta = \theta_p$. The boundstate energy is then given by
\begin{align}\label{boundstateavg}
    E_p = \frac{1}{W} \int^{W/2}_{-W/2} dy \,\, \epsilon_p (y).
\end{align}
Here $\epsilon_p(y)$ is the energy of the electron/hole which traverses the trajectory with corresponding label $(y,\theta_p)$.

\subsection*{Perfect interfaces $T_p = 1$}
To determine the characteristic energies $\epsilon_p(y)$, let us first consider the case where $T_p = 1$.
If the NS interfaces are perfectly transmitting, electron wavefunctions incident on the NS interfaces are fully Andreev reflected as holes, picking up an energy-dependent phase shift during the process.
In this case, the $\epsilon_p(y)$ is determined by the condition that the total phase accumulated throughout one cycle of a periodic trajectory $\phi_{total}$ is equal to an integer multiple of $2\pi$.

The total phase $\phi_{total}$ is given by the sum of the phase accumulated during propagation between the superconductors and the phase shifts which occur during Andreev reflection at the boundaries.
Since these contributions are gauge dependent, it will be convenient to choose a gauge where,
\begin{align}\label{gauge}
{\bf A} = By {\bf \hat x},
\end{align}
and the corresponding order parameter is given by,
\begin{align}
\Delta(x, y) = \Theta(|y| - W/2) 
\begin{cases}
\Delta e^{- i\phi/2} &  x < -\frac{L}{2}\\
0 & -\frac{L}{2} < x < \frac{L}{2} \\
\Delta e^{ i\phi/2} &  x > \frac{L}{2}\\
\end{cases}.
\end{align}
With this choice of gauge, the total phase accumulated by an electron/hole that traverses the trajectory $(y, \theta_p) $
\begin{align}
\phi_{total} =  \phi   + 2\arccos \left( \frac{\epsilon_p(y) }{\Delta} \right) + 2 L \left( \frac{\epsilon_p(y)}{v_F \cos \theta_p} \right) + 2 \Phi_m (y). 
\end{align}
The first and second terms on the RHS correspond to the phase accumulated from Andreev reflection.
The third term corresponds to the phase accumulated during free propagation between the superconductors.
The fourth term $\Phi_m (y)$ is the magnetic phase accumulated due to the presence of the vector potential.
For the choice of gauge corresponding to Eq.~\eqref{gauge}, the magnetic phase is given by,
\begin{align}\label{magneticphase}
    \Phi_m (y) = \int^{L/2}_{-L/2}  B ( y + x \tan \theta_p  )  dx = B L y,
\end{align}
For a short junction where $ L \ll \frac{\Delta}{v_F}$, the phase accumulated from free propagation is small, and the third term can be dropped. 
Enforcing the condition $\phi_{total} = 2\pi n$, we get the energy of the bound state 
\begin{align}\label{boundstateenergy}
\epsilon_p(y, \phi) = - \Delta \cos \left( \frac{\phi}{2}  + BL y \right).
\end{align}
Using Eq.~\eqref{boundstateavg}, the boundstate energy is then given by
\begin{align}\label{boundstateenergyintegral}
    E_p  = \frac{-\Delta}{W} \int^{W/2}_{-W/2} dy \,\, \cos \left( \frac{\phi}{2}  + BL y \right).
\end{align}
Changing the integration variable to $\theta = \frac{\phi}{2} + BLy$, we arrive at Eq.~\eqref{fluxAverage} for $T_p=1$
\subsection*{Imperfect interfaces $ t< 1$ }
If NS interfaces are imperfect, a portion of the incoming electron wavefunction will be scattered back via normal reflection as a hole. 
In this case, the classical trajectories no longer correspond to the periodic motion of electrons/holes, but rather to Bogoliubov quasiparticles, which are composed of electron-hole superpositions.
When a quasiparticle with incident wavefunction $\psi = (\psi_e,\psi_h)^T$ reflects off the right NS interface, both normal and Andreev reflection mix the amplitudes and phases of the reflected quasiparticle. 
When propagating across the junction, the electron and hole components of the wavefunction acquire an additional phase factor due to the magnetic field.

For the quasiparticle trajectory to be periodic, the wavefunction after scattering off the right boundary, propagating from right to left, scattering off the left boundary, and propagating from left to right, should return to its initial value.
For a trajectory with label $(y, \theta_p)$ and characteristic enregy $\epsilon_p(y)$, the condition of periodic motion is given by
\begin{align}
  \begin{pmatrix}
    \psi_e \\
    \psi_h
\end{pmatrix} 
= 
U_+ (y)
S_{L} (\epsilon_p(y) , \phi)
U_{-}(y)
S_{R} (\epsilon_p(y), \phi)
\begin{pmatrix}
    \psi_e \\
    \psi_h
\end{pmatrix} 
\end{align}
Here $S_{L,R}$ is the scattering matrix associated with the left and right interfaces,
\begin{align}
    S_{L,R} (\epsilon, \phi) =
    \begin{pmatrix}
        r(\epsilon) && t(\epsilon) e^{\mp i\phi/2}\\
        t(\epsilon) e^{\pm i\phi/2} && r(\epsilon)
    \end{pmatrix}.
\end{align}
The matrix $U_{\pm}$ describes free propagation across the junction in the $\pm x$ direction. In the short junction limit, quasiparticles traversing the junction only pick up the magnetic phase $\Phi(y)$, and the matrices are given by,
\begin{align}
    U_{\pm} (y) = 
\begin{pmatrix}
    e^{\pm i\Phi (y) } && 0 \\
    0  && e^{ \mp i\Phi(y)}
\end{pmatrix}.
\end{align}
The characteristic energy $\epsilon_p (y)$ is determined by the spectral equation,
\begin{align}\label{spectralEq}
    \mathrm{Det} \left[ 1 - 
    U_+ (y) S_{L} (\epsilon_p , \phi ) 
    U_{-}(y) S_{R}(\epsilon_p, \phi )  \right] = 0.
\end{align}
Note that $U_{\pm}$ can be absorbed into a phase shift of the scattering matrices $S_{L,R}$, allowing us to rewrite Eq.~\eqref{spectralEq} in the simplified form
\begin{align}\label{spectralEq2}
    \mathrm{Det} \left[ 1 - 
     S_{L} (\epsilon_p , \phi + 2\Phi_m ) 
     S_{R}(\epsilon_p, \phi + 2\Phi_m )  \right] = 0.
\end{align}
From here, it is clear that $\epsilon_p(y) $ can be expressed in terms of the zero field solution,
{\small
\begin{align}
    \epsilon_p (y, \phi) = \epsilon_p^{(0)} \left( \phi + 2\Phi_m (y) \right)
    = -\Delta \sqrt{1 - T_p  \sin^2 \left( \frac{\phi}{2} + \Phi_m (y) \right)}
\end{align}}
Substituting this expression into Eq.~\eqref{boundstateenergyintegral}, and changing the integration variable, we obtain Eq.~\eqref{fluxAverage}.

\section{Phase and charge basis Hamiltonian}
Discretizing $\phi$ on a uniform grid provides a direct numerical representation of the Hamiltonian, and the effect of the offset charge $n_g$ enters through the Bloch-type boundary condition
\begin{equation}
\psi(\phi+2\pi) = e^{i2\pi n_g}\,\psi(\phi).
\end{equation}
This boundary phase provides a clear physical description of charge dispersion and enables efficient diagonalization. This advantage is particularly pronounced in regimes where a charge-basis formulation would require a large number of Fourier harmonics, especially when using numerically obtained potentials that include disorder-induced perturbations from tight-binding models.

Alternatively, when the potential has a known analytic form and the Fourier series converges sufficiently rapidly, Eq.~\eqref{fluxAverage} can be expanded in Fourier harmonics, with each term modulated by a sinc envelope,
\begin{equation}
V(\phi,\Phi) 
= -\sum_{n\ge1} a_{n}(0)\;\mathrm{sinc}\!\Bigl(\tfrac{n\pi\,\Phi}{\Phi_{0}}\Bigr)\,\cos(n\,\phi)
\end{equation}
where \(\mathrm{sinc}(x)=\sin(x)/x\) and we define $ a_{n}(\Phi)=a_{n}(0)\;\mathrm{sinc}(\tfrac{n\pi\,\Phi}{\Phi_{0}})$. 
Then, the Hamiltonian \eqref{eq:FullHamiltonian} can be written in standard charge basis for numerical discretizations. 
\begin{equation}
\begin{aligned}
    &\hat{H}_{\hat{n}}(\Phi) = \sum_{n}^N 4E_C (n - n_g)^2\, |n\rangle \langle n| \\
    &\quad - \sum_{m=1}^{M} \frac{a_m(\Phi)}{2}
    \sum_{n}^N \left( |n\rangle \langle n+m| + |n+m\rangle \langle n| \right).
\end{aligned}
\end{equation}
where truncating the charge basis to $|n|\le N$ and harmonics to $m\le M <N$ controls the numerical accuracy.

\section{Tight binding model simulation}
Following Akhmerov et al.\cite{AkhermovHourglass,PhysRevB.90.155450}, we model a short and wide, two‐dimensional Josephson junction on a square lattice using the Kwant package\cite{Groth_2014}. The central scattering region of width \(W\) and length \(L\) is coupled to two superconducting leads with pairing potentials \(\Delta\) (left) and \(\Delta\,e^{i\phi}\) (right) with conventional gauge \(\mathbf{A} = (-By,0,0)\) for uniform perpendicular magnetic field. The Hamiltonian of the scattering region follows:
\begin{equation}
    H=\frac{(p-eA)^2}{2m^*}-\mu+V(x,y),
\end{equation}
where $p$ is the momentum, $e$ is the electron charge, $m^*$ is the effective mass, $\mu$ is chemical potential and $V(x,y)$ is a electrostatic potential on the scattering region. The discretization of the Hamiltonian for the tight-binding simulation reads the hopping energy $t$ with a Peierls phase accumulated from the magnetic field and the onsite energy $4t-\mu+V(x,y)$. Numerically, one constructs the scattering matrix
\begin{equation}
s \;=\; 
\begin{pmatrix}
\hat{r} & \hat{t'} \\[6pt]
\hat{t} & \hat{r'}
\end{pmatrix},
\end{equation}
from which the transmission subblock \(\hat{t}\) in the absent of flux yields transmission eigenvalues \(\{T_{p}\}\in[0,1]\) via the spectrum of \(\hat{t}\hat{t}^{\dagger}\). 
The superconducting phase difference at the boundary can be encoded via the Andreev reflection matrix
\begin{equation}
r_{A} \;=\; 
\begin{pmatrix} 
i\,e^{\,i\phi/2}\,\mathbf{I}_{N} & 0 \\[6pt]
0 & i\,e^{-\,i\phi/2}\,\mathbf{I}_{M} 
\end{pmatrix},
\end{equation}
where \(\mathbf{I}_{N}\) and \(\mathbf{I}_{M}\) are identity matrices of dimensions \(N\) (input modes) and \(M\) (output modes), respectively.  Then, one can define  
\begin{equation}
A \;\equiv\; \tfrac{1}{2}\bigl(r_{A}\,s \;-\; s^{T}\,r_{A}\bigr),
\end{equation}
and the short‐junction Andreev bound‐state condition becomes the eigenvalue problem  
\begin{equation}
\begin{pmatrix}
0 & -\,i\,A^{\dagger} \\[6pt]
i\,A & 0
\end{pmatrix}
\Psi_{\rm in} \;=\; \frac{E}{\Delta}\,\Psi_{\rm in}.
\end{equation}
The total supercurrent is then obtained
\begin{equation}
I(\phi) = \frac{2e}{\hbar} \sum_{p} \frac{dE_{p}}{d\phi},
\end{equation}
and critical current \(I_{c} = \max_{\phi} I_{J}(\phi)\) versus flux yields the Fraunhofer pattern.

For our simulations, we consider a wide and short junction geometry with a lattice constant
$a = 10\,\mathrm{nm}$, width $W = 300a$, and length $L = 10a$.
The hopping energy is $t/h = 2.3\,\mathrm{THz}$, and the chemical potential (onsite energy)
is set to $\mu/h = 5.79\,\mathrm{THz}$, with a Fermi wavelength
$\lambda_F \approx 40\,\mathrm{nm}$. In the case of imperfect boundaries, we introduce $V(x,y)=\Delta V$ for inhomogeneous chemical potential between the scattering region and superconducting leads.
To study the effect of disorder, we use a random onsite potential $V(x,y)$ uniformly distributed
in the range $[-U/2,\,U/2]$. The associated mean free path is given by \cite{AkhermovHourglass, PhysRevB.44.8017}
\begin{equation}
    l_0=\frac{6\lambda_F^3}{\pi^3a^2}\frac{\mu^2}{U^2}.
\end{equation}
In our simulations, we take $U=3\mu$, which yields a mean free path $l_0\approx13$ nm. 
\end{document}